\documentclass[twocolumn,prl,aps,showpacs,superscriptaddress,floatfix,showkeys,nofootinbib,10pt]{revtex4-2}

\usepackage{amsmath,amsfonts,amssymb,amsthm}
\usepackage{graphicx}
\usepackage{mathtools}

\newcommand{\ket}[1]{ | \, #1 \rangle} \newcommand{\bra}[1]{ \langle #1 \, |} 
\newcommand{\proj}[1]{\ket{#1}\bra{#1}}

\newcommand{\Ab}[1]{ \left| #1 \, \right|} 
\newcommand{\KullbackLeibler}[2]{D \left( #1 || #2 \right)}
\newcommand{\be}{\begin{equation}} \newcommand{\ee}{\end{equation}}
\newcommand{\ba}{\begin{aligned}} \newcommand{\ea}{\end{aligned}}

\DeclareMathOperator{\Tr}{Tr}

\DeclareMathOperator{\CNOT}{CNOT}

\DeclareMathOperator{\Pguess}{P_{guess}}

\DeclarePairedDelimiter\floor{\lfloor}{\rfloor}

\begin{document}
	
	\title{Quantum security and theory of decoherence}
	
	\author{P.~Mironowicz}
	\email{piotr.mironowicz@gmail.com}
	\affiliation{International Centre for Theory of Quantum Technologies, University of Gda\'{n}sk, Wita Stwosza 63, 80-308 Gda\'{n}sk, Poland}
	\affiliation{Department of Algorithms and System Modeling, Faculty of Electronics, Telecommunications and Informatics, Gda\'{n}sk University of Technology, Gabriela Narutowicza 11/12, 80-233 Gda\'{n}sk, Poland} 
	
	\date{\today}
	
	\begin{abstract}
		We sketch a relation between two crucial, yet independent, fields in quantum information research, \textit{viz.} quantum decoherence and quantum cryptography. We investigate here how the standard cryptographic assumption of shielded laboratory, stating that data generated by a secure quantum device remain private unless explicitly published, is disturbed by the \textit{einselection} mechanism of quantum Darwinism explaining the measurement process by interaction with the external environment.
		
		We illustrate the idea with a paradigmatic example of a quantum random number generator compromised by an analog of the Van~Eck phreaking. In particular, we derive a trade-off relation between eavesdropper's guessing probability $\Pguess$ and the collective decoherence factor $\Gamma$ of the simple form $\Pguess + \Gamma \geq 1$.
	\end{abstract}
	
	
	\maketitle
	
	Quantum cryptography~\cite{Gisin02} is one of the most spectacular successes of the quantum information theory, providing security beyond the scale accessible using classical computation techniques. Quantum devices can be used for such applications as secure key distribution~\cite{BB84} or generation of private random number~\cite{QRNG17} based on elementary physical phenomena.
	
	Still, humans can deal only with classical data, and thus at some stage, any quantum cryptographic device has to generate classical output data. This process is known as the quantum measurement~\cite{BBK95}. Even though the measurement is one of the most basic processes in quantum mechanics, it remains to be one of the most mysterious phenomena since the very beginning of the theory~\cite{Schroedinger35,Wigner63,Leggett05}. This so-called measurement problem still lacks a definitive solution, with the decoherence theory being one of the most popular approaches~\cite{Schlosshauer07}.
	
	The trailblazing works of \.Zurek~\cite{Zurek81,Zurek82} elucidated the problem in which basis the quantum measurement is actually being performed, by the introduction of the concept of the \textit{pointer basis}, \textit{~i.e.} the eigenbasis of observables commuting with the Hamiltonian determining the interaction of the measuring apparatus with the environment; the fact that the interaction with the environment is the factor that determines the measurement basis is called environment-induced superselection, or \textit{einselection}~\cite{Zurek82,Zurek03}. This result accentuated the role of the external world in the process of measurement and revealed that without this interaction, only the \textit{premeasurement}, \textit{i.e.} the correlation of the apparatus with the observed system, can occur.
	
	
	The role of the exterior world in the working of quantum devices seemingly contradicts the natural and necessary assumption, that the cryptographic devices are located inside a \textit{shielded laboratory} protecting against outflowing of the private data. Indeed, a crucial condition for the privacy of numbers, constituting e.g. a secure key, is that they remain secret unless intentionally revealed. This is particularly important for providing the information-theoretic level of security~\cite{Shannon49,Diffie76}.
	
	On the other hand, quantum Darwinism~\cite{Zurek09} suggests that without this information propagation or leakage, the decoherence will not occur, leading, in principle to the Wigner's friend paradox~\cite{Wigner61,Deutsch85}. Up to our knowledge, the problem of this prominent role of the environment has not been investigated in the context of cryptographical applications. The most related considerations concerned only the role of noise in cryptography~\cite{Brandt99,Sharma18}. This paper aims to provide an example, of how the direct connection between the low-level description of the measurement process, and the high-level specification of application protocols can be done.

	\noindent\textit{Methods.-}
	We concentrate on an elementary operation of a qubit measurement, as a basic operation for the majority of quantum devices. We follow~\cite{Zurek81} and call the observed qubit a \textit{system} $(S)$, the measuring device an \textit{apparatus} $(A)$; the third subsystem the environment $(E)$.
	
	We model the premeasurement upon rank-1 projectors $\{P_0^{(S)}, P_1^{(S)}\}$, with $P_0^{(S)}+P_1^{(S)}=\openone^{(S)}$ by a $\CNOT$ operation conditioned on them, i.e. $U^{(SA)} = P_0^{(S)} \otimes \openone^{(A)} + P_1^{(S)} \otimes \sigma_x^{(A)}$, where superscripts in parenthesis denote the subspace. The interaction of the apparatus with the environment is given by the unitary transformation: $U^{(AE)} = \proj{0}^{(A)} \otimes U_0^{(E)} + \proj{1}^{(A)} \otimes U_1^{(E)}$ leading to the decoherence in the computational basis of $(A)$.
	
	To illustrate a simple scenario of quantum randomness generation we consider measurement of the state $\ket{+}^{(S)} = 1/\sqrt{2} (\ket{0}^{(S)}+\ket{1}^{(S)})$ in the computational basis, $P_i^{(S)} = \proj{i}^{(S)}$, $i=0,1$, with (A) and (E) initially in $0$-th state leading to $\rho^{(SA)}(T) \equiv \Tr_{E} \left[ \proj{\phi}^{(SAE)} \right]$ with 
	\begin{equation}
		\ket{\phi}^{(SAE)} \equiv U^{(AE)} U^{(SA)} \ket{+}^{(S)} \otimes \ket{0}^{(A)} \otimes \ket{0}^{(E)},
	\end{equation}
	where $T$ is the time after which all the interactions occur. We get that $\Gamma = \Ab{\bra{0}^{(E)} U_1^{(E) \dagger} U_0^{(E)} \ket{0}^{(E)}}$ is the collective decoherence factor of the joint state $\rho^{(SA)}(T)$ equal to
	\begin{equation}
		\frac{1}{2}
		\begin{bmatrix}
			1 & 0 & 0 & \bra{0}^{(E)} U_1^{(E) \dagger} U_0^{(E)} \ket{0}^{(E)} \\
			0 & 0 & 0 & 0 \\
			0 & 0 & 0 & 0 \\
			\bra{0}^{(E)} U_0^{(E) \dagger} U_1^{(E)} \ket{0}^{(E)} & 0 & 0 & 1
		\end{bmatrix},
	\end{equation}
	and the full orthogonalization of measurement results occur~\cite{myPrl}. For the measurement to betide we need also the full decoherence, \textit{i.e.} $\Gamma \ll 1$.
	
	The above model refers to the simplest quantum randomness generation, where the measured state $\ket{+}^{(S)}$ is prepared in a basis that is unbiased~\cite{Bengtsson07} to the basis $\{ P_i^{(S)} \}$ in which the premeasurement is performed, and the results are stored in the computational basis of the subsystem $(A)$ that is initially preset to $\ket{0}^{(A)}$ to maximize its information capacity~\cite{Zwolak09}. The interaction part $U^{(SA)}$ is designed by the user of the quantum device that calibrates the measuring device to measure in the selected basis, possibly taking into account the characteristics of his source of states; this part is also responsible for apparatus state orthogonalization.
	
	The actual measurement is finalized by the interaction $U^{(AE)}$. That interaction is supposed to be engineered by the vendor of the measuring apparatus; the computational basis of $(A)$ is actually the one that is being \textit{displayed} to the user and the shape of $U^{(AE)}$ is determined by the device's \textit{case}, like e.g. plastic housing of a USB stick, or metal shielding of a rack-mounted multimeter.
	
	We consider a $4$-th subspace, denoted $(V)$ for Van~Eck-type eavesdropper since our approach is a quantum analog of the so-called Van~Eck attack in classical cryptography~\cite{VanEck85}, where the electromagnetic radiation of classical devices is captured by antennas and used to intercept the private content.
	
	In the attack, the eavesdropper intends to capture information regarding the measurement result stored in the apparatus. Since both the former and latter are classical data, we assume the result of wiretapping is stored in the computational basis. We consider a passive reception of the content of the environment, \textit{i.e.} eavesdropper doesn't change it. This restricts his action to a $\CNOT$ conditioned on some orthogonal projectors $\{P_0^{(E)}, P_1^{(E)}\}$, with $P_0^{(E)}+P_1^{(E)}=\openone^{(E)}$, \textit{i.e.} $U^{(EV)} = P_0^{(E)} \otimes \openone^{(V)} + P_1^{(E)} \otimes \sigma_x^{(V)}$; his initial state is $\ket{0}^{(V)}$.
	
	Direct calculations show the final state $\ket{\psi}^{(SAEV)}$ is
	\begin{equation}
		\begin{aligned}
			(&P_0^{(E)} U_0^{(E)} \ket{0000}^{(SAEV)} + P_1^{(E)} U_0^{(E)} \ket{0001}^{(SAEV)} + \\
			&P_0^{(E)} U_1^{(E)} \ket{1100}^{(SAEV)} + P_1^{(E)} U_1^{(E)} \ket{1101}^{(SAEV)})/\sqrt{2},
		\end{aligned}
	\end{equation}
	and thus the joint state of the user apparatus and the eavesdropper $\rho^{(AV)}(T) = \Tr_{SE} \left[ \proj{\psi}^{(SAEV)} \right]$ is a state diagonal in the computational basis with coefficients: $\bra{0} U_0^{\dagger} P_1 U_0 \ket{0}/2$, $\bra{0} U_0^{\dagger} P_0 U_0 \ket{0}/2$, $\bra{0} U_1^{\dagger} P_1 U_1 \ket{0}/2$, and $\bra{0} U_1^{\dagger} P_0 U_1 \ket{0}/2$, where we omitted the superscipt $(E)$.
	
	The figure of cryptographical merit we consider here is the probability that the eavesdropper correctly guesses the measurement result of the apparatus~\cite{min-entropy1,min-entropy2,min-entropy3,min-entropy4}, denoted $\Pguess$. This happens when both two-dimensional subsystems $(A)$ and $(V)$ indicate the same binary value, thus it is given by:
	\begin{equation}
		\label{eq:pguess}
		\begin{aligned}
			&\bra{00}^{(AV)} \rho^{(AV)}(T) \ket{00}^{(AV)} + \bra{11}^{(AV)} \rho^{(AV)}(T) \ket{11}^{(AV)} \\
			&\ = \frac{1}{2} \bra{0}^{(E)} (U_0^{(E) \dagger} P_0^{(E)} U_0^{(E)} + U_1^{(E) \dagger} P_1^{(E)} U_1^{(E)}) \ket{0}^{(E)}.
		\end{aligned}
	\end{equation}
	The environment $(E)$ mediates between subsystems $(A)$ and $(V)$, and can be of much larger dimension.
	
	From~\eqref{eq:pguess} we see, that the guessing probability depends both on the ability of the environment to gather information regarding the apparatus, modeled by $\{U_0^{(E)}, U_1^{(E)}\}$, and on the possibility of collecting the signal from the environment through the Van~Eck-type antenna, modeled by $\{P_0^{(E)}, P_1^{(E)}\}$. The shielded laboratory assumption refers to the case with $U_0^{(E)} = U_1^{(E)} = \openone^{(E)}$; then the value of~\eqref{eq:pguess} is $0.5$, so no information leaks outside the laboratory, and simultaneously $\Gamma = 1$, thus the measurement does not occur.
	
	The shielding determines $U^{(AE)}$ and is dependent on the owner of the laboratory (and the technology used) and should be considered as a part of the quantum device. The antenna determines $U^{(EV)}$ and is possessed by the wiretapper, and its capabilities are limited by his resources, reflecting his control over the information scattering. The rest of this paper aims to model the dependence of the guessing probability~\eqref{eq:pguess} on the power of the eavesdropper.

	\noindent\textit{Results.-}
	Now, let us use the above results to analyse a case of the environment consisting of $N$ qubits. We follow the standard approach~\cite{Zurek21,Mironowicz22} and model the $U^{(AE)}$ interaction as $N$ independent imperfect $\CNOT$ defined as $U_{\oslash}(\theta) \equiv \proj{0}^{(A)} \otimes \openone_2^{(\cdot)} + \proj{1}^{(A)} \otimes P_{\oslash}^{(\cdot)}$, with $P_{\oslash} \equiv \begin{pmatrix} \sin \theta & \cos \theta \\ \cos \theta & -\sin \theta \end{pmatrix}$, and $\theta$ fixed for the setup. Thus, we have $U_0^{(E)} = \openone_{2^N}$ and $U_1^{(E)} = \bigotimes_{s=1}^N P_{\oslash}^{(s)}$, where $(s)$ denotes $s$-th environmental qubit.
		
	From this it follows that the collective decoherence factor $\Gamma = \Ab{\sin \theta}^N$, or, that for a specific value of $\Gamma$ an interaction with at least $N \geq \frac{-\ln \Gamma}{-\ln \Ab{\sin \theta}}$ environmental qubits is required. The factor dependent on $\theta$ is an engineering parameter, and $\Gamma$ is a quantumness parameter, thus we may assume that the number $n \leq N$ of qubits accessible to the eavesdropper is $\frac{\mu(-\ln \Gamma)}{-\ln \Ab{\sin \theta}}$, for some function $\mu$.
	
	Let us consider the case when the eavesdropper is not able to perform a coherent measurement on multiple qubits, and needs to perform the guess basing on many separate single-qubit measurements. If he performs the Helstrom measurement~\cite{Helstrom69}, with one of the projectors given by $\begin{pmatrix} \cos^2(\theta/2) & -(\sin \theta) /2 \\ -(\sin \theta) /2 & \sin^2(\theta/2) \end{pmatrix}$, on a specific environmental qubit, the success probability of correct distiguishing its state is $p \equiv (1 + \Ab{\cos \theta}) / 2$.
	
	Suppose that the guess is given as the majority of $n$ single-qubit guesses, \textit{i.e.} it succeeds when at least $n/2$ of these guesses is correct. Thus, the total success probability of the guess~\eqref{eq:pguess} is equal to $1-F(n/2;n,p)$, where $F(\cdot;n,p)$ is the cumulative distribution function (CDF) of the binomial distribution with $n$ Bernoulli trials with success probability $p$. Now, we ask, for what range of $\Gamma$, $\theta$, and $\mu$ do we have $\Pguess \gg 1/2$?
	
	It can be shown~\cite{Bernstein64,Ferrante21} that for $p \in (0,1)$ and $a < p$ it holds $F(a n;n,p) \leq \exp \left( -n \KullbackLeibler{a}{p} \right)$, where $\KullbackLeibler{a}{p} \equiv a \ln{\frac{a}{p}} + (1-a) \ln{\frac{1-a}{1-p}}$ is the Kullback–Leibler divergence between Bernoulli random variables. Using the above formulae for $p$ and $n$ we directly get $\KullbackLeibler{1/2}{(1 + \Ab{\cos \theta}) / 2} = -\ln \Ab{\sin \theta}$, and so
	\begin{equation}
		\label{eq:Pguess}
		\Pguess \geq 1 - \exp \left( - \mu(-\ln \Gamma)\right),
	\end{equation}
	and the lower bound doesn't depend on $\theta$. From these considerations it follows that taking any $\mu$ satisfying $\lim_{x \to \infty} \mu(x) = \infty$ and $\lim_{x \to \infty} \frac{\mu(x)}{x} = 0$ we have that in the classical limit an arbitrary small fraction of all environmental qubits is enough to provide the eavesdropper full access to cryptographic data.
	
	This complies with the \textit{information plateau} observation of the quantum Darwinism~\cite{Zurek09}. We also note when the whole environment is accessible to the eavesdropper, even in incoherent, semi-classical, manner, \textit{i.e.} for $\mu(x) = x$, the relation~\eqref{eq:Pguess} takes a simple trade-off form 
	\begin{equation}
		\label{eq:tradeoff}
		\Pguess + \Gamma \geq 1.
	\end{equation}
	We see that the shielded laboratory assumption $\Pguess \leq 1/2$ entails $\Gamma \geq 1/2$, \textit{viz.} restricts the measurement to premeasurement. The trade-off~\eqref{eq:tradeoff} in particular states that the eavesdropper’s ability to read out the information of the measurement limits the degree of decoherence.
	
	We note that the above model is exceedingly simplistic, covering only a particular form of potential attacks of Van~Eck's type, and may not be the most efficient one. Yet, this restricted and fairly simple and natural form of gathering information from the surroundings is enough to compromise the security of a device producing private numbers showing that the discussed sort of attacks is a serious threat.
	
	Let us summarize the assumptions we make in the derivation of the trade-off~\eqref{eq:tradeoff}. We assume a particular form of the interaction $U^{(SA)}$ justified by the functioning of a measuring device. The decomposition of the measurement process into parts $U^{(SA)}$ and $U^{(AE)}$ is justified by its logical order in the measurement, \textit{i.e.} first occurs the premeasurement, and then occurs the decoherence. Thus, stating that the measuring device interacts with the environment \textit{via} some interaction $U^{(AE)}$ is not restrictive. Next, we perform the calculations using a particular form of $U^{(AE)}$ used in~\cite{Zurek21,Mironowicz22}; we leave considerations with more general $U^{(AE)}$ as an important new engineering task of designing cryptographical devices in a way more secure against Van~Eck's attacks. The considered form of the interaction $U^{(EV)}$ doesn't restrict the generality of our results, as it is sufficient for the trade-off relation to occur. We show that such interaction exists, possibly there exists another interaction $U^{(EV)}$ for which the trade-off relation is even tighter; we also leave this for a further study of the interplay between designing devices with more suitable $U^{(AE)}$ and attacks with more efficient $U^{(EV)}$.
	
	To see consequences of the above analysis, we start with the simplest case with one environmental qubit interacting \textit{via} a perfect CNOT, \textit{viz.} $N = 1$ and $\theta = 0$. We have the full decoherence with $\Gamma = 0$ but, if the only environmental qubit is intercepted by the eavesdropper, we also have $\Pguess = 1$. For a toy model of decoherence with $N = 20$ and $\theta = \pi / 4$ we have $\Gamma \approx 0.001$; then for $1$, $3$, and $5$ intercepted environmental qubits $\Pguess$ is $0.85$, $0.94$, and $0.98$, respectively. For the more realistic case with $\Gamma \approx 10^{-40}$~\cite{Zurek86} if the van~Eck's antenna observes $1\%$ or $5\%$ of the environment, then $\Pguess$ is $0.6$ or $0.99$, respectively.
	
	To investigate how the privacy of quantum random numbers from the above model is compromised by a coherent Van~Eck-type antenna we performed also numerical simulations. Let $D^{(E)}$ denote the dimension of the environment, and $k \in \{2, \cdots, D^{(E)}\}$ be the number of degrees of freedom of the environment the antenna can faithfully distinguish; the ratio $k/D^{(E)}$ can be considered as the measure of how much of the environment is monitored, or controlled, by the eavesdropper.
	
	In the numerical calculations we consider Haar distributed~\cite{Haar33} $\{U_0^{(E)}, U_1^{(E)}\}$. To simulate the limitations on the Van~Eck's antenna we now consider $P_0^{(E)}$ of the following form. We decompose the space of the environment $\mathcal{H}^{(E)}$ into two parts: $\mathcal{H}^{(\hat{E})}$, and $\mathcal{H}^{(\tilde{E})}$, with dimensions $k$ and $D^{(E)} - k$, respectively; so that $\mathcal{H}^{(E)} = \mathcal{H}^{(\hat{E})} \oplus \mathcal{H}^{(\tilde{E})}$, where $\oplus$ denotes the direct sum of spaces. We take $P_0^{(E)} = P^{(\hat{E})} \oplus \openone^{(\tilde{E})}$, where $P^{(\hat{E})}$ is an arbitrary projector on $\mathcal{H}^{(\hat{E})}$ with the rank $\floor{k/2}$.
	
\begin{figure}
		\centering
		\includegraphics[width=0.95\linewidth]{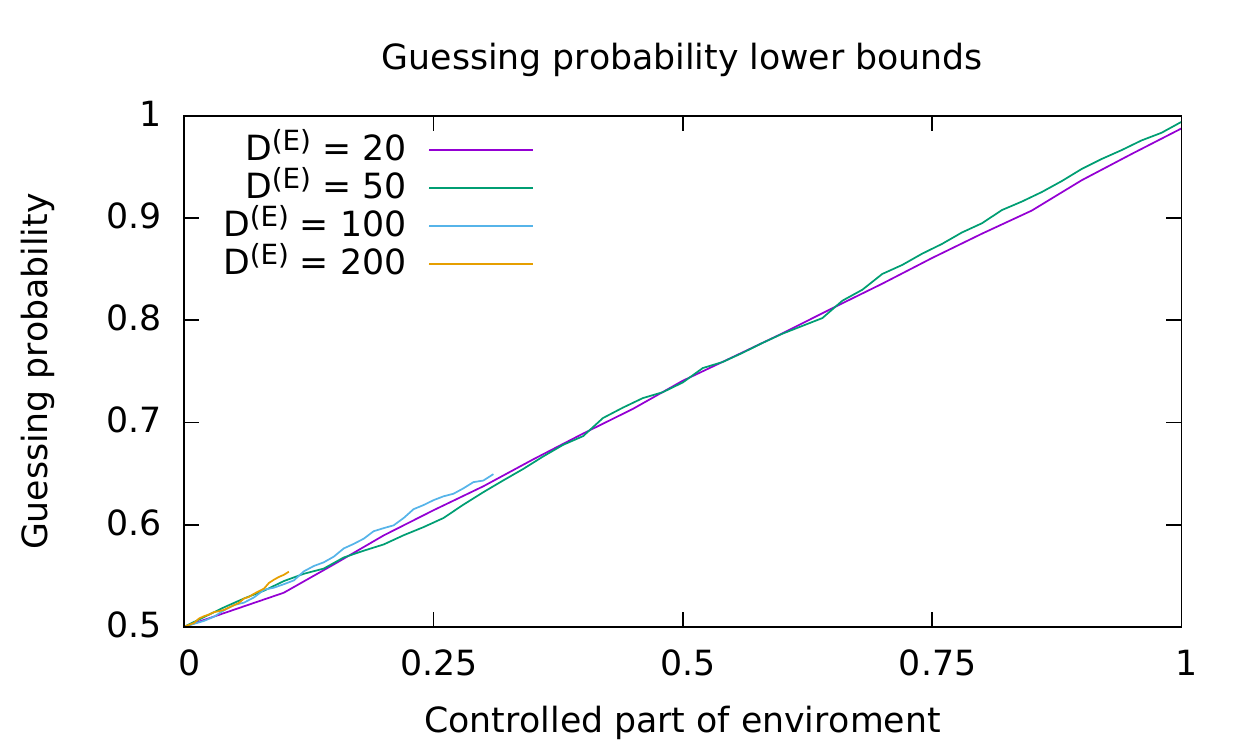}
		\caption{The dependence of the guessing probability of the indicator of the measuring apparatus by a Van~Eck-type eavesdropper depending on the relative size of the part of the environment observed by the wiretapping antenna for different dimensions $D^{(E)}$ of the environment.}
		\label{fig:pguess}
	\end{figure}

	We executed the computation of~\eqref{eq:pguess} for $D^{(E)} = 20, 50, 100, 200$. To this end, for each instance we parametrized the operator $P^{(\hat{E})}$ and performed gradient search to maximize the value of the quessing probability. We averaged the results of several ($15$, $8$, $11$, and $4$, respectively) instances with different $\{U_0^{(E)}, U_1^{(E)}\}$. The results are shown at Fig.~\ref{fig:pguess}.
	
	It can be observed that the guessing probability is more or less proportional to the observed part of the environment. We note that even when the eavesdropper possesses full access to the environment's information, he still may not be able to achieve the value $1$ of guessing probability since not all information could have been propagated, especially when the value of $D^{(E)}$ is small. This relates to the situation with $\Gamma \gg 0$, so with no full measurement inside the laboratory.

	\noindent\textit{Conclusions.-}
	Despite this work being embedded in the framework of einselection and quantum Darwinism, we don't consider here the usual scenario of information widespread in multiple copies of independent parts of the environment. We concentrate on the observation of a single observer, so, this cannot be understood as a model of \textit{objectivity}~\cite{Korbicz21} (or \textit{inter-subjectivity}~\cite{myPrl,Ajdukiewicz78}) as investigated in the recent works~\cite{Zurek21}. Yet, it is obvious, that after a measurement is performed, then knowing \textit{what} has been measured (\textit{i.e.} the basis), should imply the ability to copy and disseminate the result~\cite{Wootters82,Zurek00}.
	
	We have seen that the shielded laboratory assumption prevents the occurrence of measurement; and that by relaxing this assumption, we open a way for attacks similar to the Van~Eck phreaking. We note that although Fig.~\ref{fig:pguess} shows cases with relatively small sizes of the environment compared to macroscopic objects, it suggests that the greater the dimension, the lower part of the laboratory's surroundings has to be under control for the significant potential for eavesdropping. Indeed, the relation~\eqref{eq:tradeoff} we derived for incoherent qubits phreaking clearly indicates that any sort of cryptographic protocol is prone to the discussed type of attacks. We would like to stress that our topic is not an analysis of the case when the device that processes quantum information happens not to be perfectly shielded due to imperfections; on contrary, we show that for any quantum measuring-based device to function properly it is \textit{necessary to drop} the perfect shielding assumption by the design.
	
	In this preliminary study, we investigated only the simplest case where the quantum randomness is obtained from the measurement on a different basis than the prepared state. Although simple, this scenario is ubiquitous as an ingredient of more involved and complex quantum protocols.
	
	This work intends to show that the quantitative investigation of the relation between two important, yet till now disjoint, areas of quantum information, \textit{viz.} theory (quantum Darwinism) and application (quantum cryptography) of measurements, is possible. Quantum cryptography is a wide field and is currently the only quantum information research area with serious commercial deployments~\cite{Hejamadi18}. Our main premise is to change one of the essential parts of the paradigm of quantum cryptography that was based on neglecting, or abstracting from, the way the quantum measurement is performed in cryptographic devices.
	
	We expect the presented result will encourage researchers working on decoherence theory to contribute to the development of the design of cryptographic devices, similarly as they contribute to the area of quantum computation~\cite{Zurek95,Zurek18}. We consider it an interesting and vital problem, how such analysis can be extended to more complicated scenarios, and cover such problems as quantum communication~\cite{BB84} or quantum key distribution~\cite{Renner08}, not only in a device-dependent scenario, like in this work, but possibly in device-independent~\cite{MY98}, or semi-device-independent~\cite{Marcin11} frameworks.
	
	We close this work with the practical open question of whether it is possible to protect against the introduced type of attacks? We predict the general answer, with the eavesdropper with sufficient control over the environment, to be negative. Still, it is plausible that under some reasonable assumptions regarding the technology of the eavesdropper, one can engineer the shielding in such a way that the measurement does occur while the wiretapping task becomes burdensome.

	\noindent\textit{Acknowledgments.-}
	The work is supported by the Foundation for Polish Science (IRAP project, ICTQT, contract no. 2018/MAB/5, co-financed by EU within Smart Growth Operational Programme) and NCBiR QUANTERA/2/2020 (www.quantera.eu)	under the project eDICT. The numerical calculations we conducted using OCTAVE~6.1~\cite{OCTAVE}, and packages QETLAB~0.9~\cite{QETLAB} and Quantinf~0.5.1~\cite{quantinf}.

	

\begin{thebibliography}{99}
		
		\bibitem{Gisin02} Gisin, N., Ribordy, G., Tittel, W., Zbinden, H., \textit{Quantum cryptography}, Reviews of Modern Physics, 74(1), 145 (2002).
		\bibitem{BB84} Bennett, C. H., Brassard, G., \textit{Quantum cryptography: Public key distribution and coin tossing}, Proceedings of the International Conference on Computers, Systems and Signal Processing, Bangalore, India, pp. 175-179, arXiv:2003.06557 (1984).
		\bibitem{QRNG17} Herrero-Collantes, M., Garcia-Escartin, J. C., \textit{Quantum random number generators}, Reviews of Modern Physics, 89(1), 015004 (2017).
		\bibitem{BBK95} Braginsky, V. B., Braginskiĭ, V. B., Khalili, F. Y., \textit{Quantum measurement}, Cambridge University Press (1995).
		\bibitem{Schroedinger35} Schroedinger, E., \textit{Die gegenwärtige Situation in der Quantenmechanik}, Naturwissenschaften, 23(49), 823-828 (1935).
		\bibitem{Wigner63} Wigner, E. P., \textit{The problem of measurement}, American Journal of Physics, 31(1), 6-15 (1963).
		\bibitem{Leggett05} Leggett, A. J., \textit{The quantum measurement problem}, Science, 307(5711), 871-872 (2005).
		\bibitem{Schlosshauer07} Schlosshauer, M. A., \textit{Decoherence: and the quantum-to-classical transition}, Springer Science and Business Media (2007).
		\bibitem{Zurek81} \.Zurek, W. H., \textit{Pointer basis of quantum apparatus: Into what mixture does the wave packet collapse?}, Physical Review D, 24(6), 1516 (1981).
		\bibitem{Zurek82} \.Zurek, W. H., \textit{Environment-induced superselection rules}, Physical Review D, 26(8), 1862 (1982).
		\bibitem{Zurek03} \.Zurek, W. H., \textit{Decoherence, einselection, and the quantum origins of the classical}, Reviews of Modern Physics, 75(3), 715 (2003).
		\bibitem{Shannon49} Shannon, C. E., \textit{Communication theory of secrecy systems}, The Bell system technical journal, 28(4), 656-715 (1949).
		\bibitem{Diffie76} Diffie, W., Hellman, M., \textit{New directions in cryptography}, IEEE transactions on Information Theory, 22(6), 644-654 (1976).
		\bibitem{Zurek09} \.Zurek, W. H., \textit{Quantum Darwinism}. Nature physics, 5(3), 181-188 (2009).
		\bibitem{Wigner61} Wigner, E. P., \textit{Remarks on the mind-body question}, In I. J. Good (ed.), The Scientist Speculates. Heineman (1961).
		\bibitem{Deutsch85} Deutsch, D., \textit{Quantum theory as a universal physical theory}, International Journal of Theoretical Physics, 24(1), 1-41 (1985).
		\bibitem{Brandt99} Brandt, H. E., \textit{Qubit devices and the issue of quantum decoherence}, Progress in Quantum Electronics, 22(5-6), 257-370 (1999).
		\bibitem{Sharma18} Sharma, V., Shrikant, U., Srikanth, R., Banerjee, S., \textit{Decoherence can help quantum cryptographic security}, Quantum Information Processing, 17(8), 1-16 (2018).
		\bibitem{Bengtsson07} Bengtsson, I., \textit{Three ways to look at mutually unbiased bases}, In AIP Conference Proceedings (Vol. 889, No. 1, pp. 40-51). American Institute of Physics (2007).
		\bibitem{Zwolak09} Zwolak, M., Quan, H. T., \.Zurek, W. H., \textit{Quantum Darwinism in a mixed environment}, Physical Review Letters, 103(11), 110402 (2009).
		\bibitem{VanEck85} Van~Eck, W., \textit{Electromagnetic radiation from video display units: An eavesdropping risk?}, Computers and Security, 4(4), 269-286  (1985).
		\bibitem{min-entropy1} Chor, B., Goldreich, O., \textit{Unbiased bits from sources of weak randomness and probabilistic communication complexity}, SIAM Journal on Computing, 17(2), 230-261 (1988).
		\bibitem{min-entropy2} Impagliazzo, R., Levin, L. A., Luby, M., \textit{Pseudo-random generation from one-way functions}, In Proceedings of the twenty-first annual ACM symposium on Theory of computing (pp. 12-24) (1989).
		\bibitem{min-entropy3} Konig, R., Renner, R., Schaffner, C., \textit{The operational meaning of min-and max-entropy}, IEEE Transactions on Information theory, 55(9), 4337-4347 (2009).
		\bibitem{min-entropy4} Issa, I., Wagner, A. B., \textit{Measuring secrecy by the probability of a successful guess}, IEEE Transactions on Information Theory, 63(6), 3783-3803 (2017).
		\bibitem{Zurek21} Touil, A., Yan, B., Girolami, D., Deffner, S., \.Zurek, W. H., \textit{Eavesdropping on the Decohering Environment: Quantum Darwinism, Amplification, and the Origin of Objective Classical Reality}, Physical Review Letters 128(1), 010401 (2022).
		\bibitem{Mironowicz22} Mironowicz, P., Horodecki, P., Horodecki, R., \textit{Non-Perfect Propagation of Information to a Noisy Environment with Self-Evolution}, Entropy 24(4), 467. (2022).
		\bibitem{Helstrom69} Helstrom, C. W. Quantum detection and estimation theory. Journal of Statistical Physics, 1(2), 231-252 (1969).
		\bibitem{Bernstein64} Bernstein, S. N., \textit{Collected works}, vol. 4. Izdat. Akad. Nauk SSSR, Moscow (1964).
		\bibitem{Ferrante21} Ferrante, G. C., \textit{Bounds on Binomial Tails With Applications}, IEEE Transactions on Information Theory, 67(12), 8273-8279 (2021).
		\bibitem{Zurek86} \.Zurek, W. H., \textit{Reduction of the wavepacket: How long does it take?}, In Frontiers of Nonequilibrium Statistical Physics (pp. 145-149), Springer, Boston, MA (1986).
		\bibitem{Haar33} Haar, A., \textit{Der Massbegriff in der Theorie der kontinuierlichen Gruppen}, Annals of mathematics, 147-169 (1933).
		\bibitem{Korbicz21} Korbicz, J. K., \textit{Roads to objectivity: Quantum Darwinism, Spectrum Broadcast Structures, and Strong quantum Darwinism – A review}, Quantum, 5, 571 (2021).
		\bibitem{myPrl} Mironowicz, P., Korbicz, J. K., Horodecki, P., \textit{Monitoring of the process of system information broadcasting in time}, Phys. Rev. Lett. \textbf{118}, 150501 (2017).
		\bibitem{Ajdukiewicz78} Ajdukiewicz, K., Giedymin, J., \textit{The scientific world-perspective and other essays}, 1931-1963 (pp. 155-164), Dordrecht: Reidel (1978).
		\bibitem{Wootters82} Wootters, W. K., \.Zurek, W. H. \textit{A single quantum cannot be cloned}, Nature, 299(5886), 802-803 (1982).
		\bibitem{Zurek00} \.Zurek, W. H.. \textit{Schrödinger's sheep}, Nature, 404(6774), 130-131 (2000).
		\bibitem{Hejamadi18} Shenoy-Hejamadi, A., Pathak, A., Radhakrishna, S., \textit{Quantum cryptography: key distribution and beyond}, Quanta, 6(1), 1-47 (2018).
		\bibitem{Zurek95} Chuang, I. L., Laflamme, R., Shor, P. W., \.Zurek, W. H., \textit{Quantum computers, factoring, and decoherence}, Science, 270(5242), 1633-1635 (1995).
		\bibitem{Zurek18} Gardas, B., Dziarmaga, J., \.Zurek, W. H., Zwolak, M., \textit{Defects in quantum computers}, Scientific Reports, 8(1), 1-10 (2018).
		\bibitem{Renner08} Renner, R., \textit{Security of quantum key distribution}, International Journal of Quantum Information, 6(01), 1-127 (2008).
		\bibitem{MY98} Mayers, D., Yao, A., \textit{Quantum cryptography with imperfect apparatus}, In Proceedings 39th Annual Symposium on Foundations of Computer Science (Cat. No. 98CB36280) (pp. 503-509). IEEE (1998).
		\bibitem{Marcin11} Paw\l{}owski, M., Brunner, N., \textit{Semi-device-independent security of one-way quantum key distribution}, Physical Review A, 84(1), 010302(R) (2011).
		\bibitem{OCTAVE} Eaton, J. W., Bateman, D., Hauberg, S., Wehbring, R., \textit{{GNU Octave} version 6.1.0 manual: a high-level interactive language for numerical computations}, \url{https://www.gnu.org/software/octave/doc/v6.1.0/} (2020).
		\bibitem{QETLAB} Johnston, N., \textit{{QETLAB}: A {MATLAB} toolbox for quantum entanglement, version 0.9}, \url{http://qetlab.com} (2016).
		\bibitem{quantinf} Toby Cubitt, Quantinf Matlab Package, version 0.5.1, \url{https://www.dr-qubit.org/matlab.html} (2013).
		
	\end{thebibliography}
\end{document}